\documentclass[%
12pt,T
 oneocolumn,
 nofootinbib,
 amsmath,amssymb,
 aps,
floatfix,
unsortedaddress
]{revtex4-2}

\usepackage{graphicx,color}
\usepackage{subfig}
\usepackage{dcolumn}
\usepackage{bm}
\usepackage{graphicx,color}
\usepackage{float}

\usepackage{amsmath}
\usepackage{mathtools}

\usepackage{cancel}
\usepackage[pdfpagelabels, pdfencoding=auto, psdextra]{hyperref}
\hypersetup{%
 pdfsubject=Paper,
 pdfkeywords={nuclear physics} {Pionless EFT} {Two-Nucleon} {Lattice} {L\"usher formula},
 unicode = true,
 breaklinks = true,
 colorlinks = true,
 linkcolor = blue,
 citecolor = blue,
 menucolor = blue,
 citecolor = blue,
 urlcolor = blue
}

\graphicspath{{./}}
\begin{document}

\title{Examination of the  $ \phi-NN $ bound-state problem with lattice QCD $ N-\phi $  potentials
}
\author{Faisal Etminan}
 \email{fetminan@birjand.ac.ir}
\affiliation{
 Department of Physics, Faculty of Sciences, University of Birjand, Birjand 97175-615, Iran
}%
\affiliation{ Interdisciplinary Theoretical and Mathematical Sciences Program (iTHEMS), RIKEN, Wako 351-0198, Japan}

\author{Amanullah Aalimi}
 \email{amanullahaalimi33@birjand.ac.ir}
\affiliation{
Department of Physics, Faculty of Sciences, University of Birjand, Birjand 97175-615, Iran
}%

\date{\today}%

\begin{abstract}
			The developed Faddeev three-body equations are solved to search for bound-state solutions of a phi-meson $ \left(\phi\right) $ and two nucleons $ \left( NN \right) $ system. The newly  published spin $ 3/2 $  $ N-\phi $ potential based on the $ \left({2+1}\right) $-flavor lattice QCD simulations  near the physical point, and the realistic $ NN $  Malfliet-Tjon (MT) potential, are employed. Our numerical calculations for $ (I)J^{\pi}=(0)2^{-}$ $ \phi{-d} $ system in maximum spin  lead to ground state binding energy of about $ 7 $  MeV and a matter radius of about $ 8 $ fm. Our results indicate the possibility of the formation of new nuclear clusters.
\end{abstract}


\maketitle

\section{Introduction} \label{sec:intro}
In strangeness nuclear physics, the possibility of the formation of $ \phi $-mesic bound states with nucleons (N)~\cite{PhysRevC.95.055202,HIRENZAKI2010406,10.1143/PTP.124.147,PhysRevC.96.035201} is one the most exciting and important fields due to the quark content of the $  \phi $-meson as being $ s\bar{s} $. One of the simplest candidates for $ \phi $-mesic nuclei can be the $ \phi{-NN} $ {system}~\cite{Belyaev2008,Belyaev2009,Sofianos2010,PhysRevC.108.034614}. Accordingly, the binding energy of $\phi{-NN}$  state  has been calculated using the folding method~\cite{Belyaev2008},  solving the Faddeev equations in the coordinate space ~\cite{Belyaev2009}  and in two-variable integro-differential equations on the $ {D=}{3}\left({A-1}\right) $-dimensional space~\cite{Sofianos2010}, 
where mostly attractive phenomenological  $ {N-}\phi $ interaction~\cite{PhysRevC.63.022201} by  central binding energy of $ 9.47  $ MeV and the semi-realistic Malfliet-Tjon (MT) $ {NN} $ potential\cite{MALFLIET1969161}, are employed. 
They concluded that the $\phi{-NN}$ system is bound with a binding energy of about $ 40\left(23\right) $ MeV for triplet(singlet) $ {NN} $ interaction~\cite{Sofianos2010}.

On the other hand, from the experimental point of view,  ALICE
collaboration measured the correlation function of proton-$\phi$  in heavy-ion collisions  ~\cite{PhysRevLett.127.172301}, 
together by indicating a $ p{-}\phi $ bound state using two-particle
correlation functions~\cite{CHIZZALI2024138358} {request} the proton${-}\phi$ bound state hypothesis.
To go one step further, the femtoscopic analysis of hadron-deuteron ($ hd $) correlation functions
could play a crucial role in understanding the structure and
dynamics of the atomic nuclei. Therefore the three-body calculations of the $ hd $ correlation function
{have} special importance~\cite{PhysRevC.108.064002,alicecollaboration2023exploring}.
Accordingly, the $ hd $ correlation functions are investigated for some hadron-deuteron systems like
$ pd $   \cite{PhysRevC.108.064002,mrowczynski2019hadron}, $ {K}^{-}d $ \cite{alicecollaboration2023exploring,mrowczynski2019hadron},
$ \Lambda d $ \cite{Haidenbauerprc}, $ \Xi d $ \cite{Ogata2021},   and the production of $ \Omega NN $ 
in ultra-relativistic heavy ion collisions  {are}  studied in~\cite{zhang2021production,ETMINAN2023122639}.
Hence, it is more desirable to study the properties of $\phi{-d}$  system by using the state-of-the-art few-body computational method .

Very recently,  the first   lattice QCD {simulations} of the ${N-}\phi$ potential
in $^{4}S_{3/2}$ channel (by the notation $^{2s+1}L_{J}$ where $s, L$ and
$J$ are the total spin, orbital angular momentum and total angular momentum,
respectively) are published~\cite{yan2022prd}. The simulation is performed by $\left(2+1\right)$-flavor
with quark masses near the physical point $m_{\pi}\simeq146.4$ MeV
and $m_{K}\simeq525$ MeV on a sufficient large lattice size of 
$\simeq\left(8.1\:{fm}\right)^{3}$. 
Here, we employ this ${N-}\phi$ potential to study $\phi{-d}$ in the
highest spin $\left(I\right)J^{p}=\left(0\right)2^{-}$, this channel
is selected because it cannot couple to the lower channels $\Lambda KN$ and
$\Sigma KN$ with the $\Lambda K$ and $\Sigma K$ subsystem in $ D $
waves. In addition, because of the small phase space, the decay to
final states by four or more particles like $\Sigma\pi KN,\Lambda\pi KN$ and
$ \Lambda\pi\pi KN $ are supposed to be suppressed~\cite{yan2022prd}. 

 We should emphasize that since, some coupling constants like $ \phi \rho \pi $ are
known to be far from being OZI (Okubo-Zweig-Iizuka ) suppressed, the coupling to channels
like $ \rho N $ or $ \pi \Delta $ could be sizable and, this has been all ignored
in the lattice simulations~\cite{yan2022prd}. Such a coupling would change the scenario that will be
studied here considerably.

Motivated by the above discussion, the binding energies and  matter radius of $ \phi{-d} \left(0\right)2^{-} $  state  in cluster model based on expansion in hyperspherical harmonics (HH method) \cite{Zhukov93,Casal2020,ETMINAN2023122639} are calculated in this paper. For $ {NN} $ interaction we have considered the semi-realistic Malfliet-Tjon $ {NN} $ potential.

The paper is organized as follows. In Section~\ref{sec:Three-body hyperspherical}  a brief sketch of the three-body hyperspherical basis formalism is given.
The  input two-body potentials  are described in Section ~\ref{sec:Two-body-potentials}.
In Section~\ref{result}, the numerical results are discussed. And the last Section~\ref{sec:Summary-and-conclusions}, is devoted to a summary and conclusion.

\section{ Three-Body Bound State by Expansion on Hyperspherical Harmonics Method }
\label{sec:Three-body hyperspherical}
The expansion on hyperspherical harmonics method is a developed  version of the Faddeev equations in coordinate space to study weakly bound three-body systems. The method is well-described in~\cite{Zhukov93,raynal1970,face}, so we present  an outline of it very briefly here. 

\begin{figure*}[hbt!]
	\centering
	\includegraphics[scale=0.3]{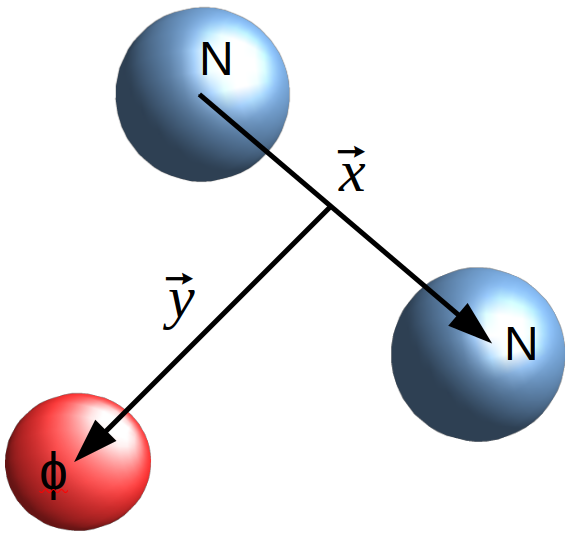}
	\caption{The Jacobi T-coordinate system used
		to describe the $ \phi{-NN} $ system. It is  noteworthy that there are three different Jacobi systems.}
	\label{fig:T_jacobi}
\end{figure*}

The complete  three-body wave function $\Psi$ is sum of three components
$\Psi^{(i)}$, i.e, $\Psi=\sum_{i=1}^{3}\Psi^{\left(i\right)}$. The components
$\Psi^{\left(i\right)}$ are function of the three different sets of
Jacobi coordinates (One of the three sets is shown in the Fig.~\ref{fig:T_jacobi}), and they satisfy the three Faddeev coupled equations,
\begin{equation}
	\left(T-E\right)\Psi^{\left(i\right)}+V_{jk}\left(\Psi^{\left(i\right)}+\Psi^{\left(j\right)}+\Psi^{\left(k\right)}\right)=0,
	\label{eq:faddeev_coupled-eq}
\end{equation}
where $T$ is the kinetic energy, $E$ is the total energy, $V_{jk}$ is
the two-body interactions between the corresponding pair and the indexes
$i,j,k$ is a cyclic permutation of $\left(1,2,3\right)$.

The Jacobi coordinates $ \left\{ \vec{x},\vec{y}\right\} $, as depicted in Fig.~\ref{fig:T_jacobi}, are employed to define the framework of three-body systems.
The Jacobi-T coordinate set is the most suitable one for the  $ \phi{-NN} $ 
system because the antisymmetrization of the wave function should be preserved 
under exchange of nucleons linked by the $ \vec{x} $
coordinate.
The variable $ \vec{x} $ represents the relative coordinates between two of the particles and $ \vec{y} $ is between their center  of mass and the third particle, both with a scaling mass factor.
From the Jacobi coordinates, we can define the hyperspherical coordinates $\{ \rho, \alpha,\Omega_{x},\Omega_{y} \} $, with hyperradius (generalized radial coordinate) $\rho^{2}=x^{2}+y^{2}$ and the hyperangle (generalized angle) $\alpha=\arctan (x/y )$. The $ \Omega_{x} $ and $ \Omega_{y} $ are the angles defining the direction of $ \vec{x} $ and $ \vec{y} $, respectively. For the sake of simplicity, we describe all angular dependencies by $\phi = (\alpha,\Omega_{x},\Omega_{y} )$.

The Hamiltonian of a three-body system in hyperspherical coordinates is defined as follows:
\begin{equation}
	\hat{H}=\hat{T}\left(\rho,\phi\right)+\hat{V}\left(\rho,\phi\right), \label{eq:Hamiltonian}
\end{equation}
the $\hat{V}(\rho,\phi)$ is potential operator which is the summation of pair interactions
and the $ \hat{T} (\rho,\phi ) $  is the free Hamiltonian operator~\cite{lay2012},
\begin{equation}
	\hat{T} (\rho,\phi )=-\frac{\hbar^{2}}{2m} \left ( \frac{\partial^{2}}{\partial\rho^{2}}+\frac{5}{\rho}\frac{\partial}{\partial\rho}-\frac{1}{\rho^{2}}\hat{K}^{2} (\phi ) \right ),
\end{equation}
where $\hat{K}$ is hyperangular momentum (generalized angular momentum) operator and $ m $ is a normalization mass for which we choose $ m = m_{N} $. 

Employing the hyperspherical coordinates, the solutions of the Schr\"odinger equation with the three-body Hamiltonian of Eq.~\eqref{eq:Hamiltonian} by total angular momentum $ j $ can be expanded for each $ \rho $ as

\begin{equation}
	\psi_{i\beta}^{j\mu}(\rho,\phi)=R_{i\beta}(\rho)\mathcal{Y}_{\beta}^{j\mu}(\phi),
\end{equation}
where the functions $\mathcal{Y}_{\beta}^{j\mu}(\phi)$ are a complete
set of hyperangular  functions 
that can be expanded in hyperspherical harmonics~\cite{Zhukov93,Casal2020,raynal1970},
and $R_{i\beta}\left(\rho\right)$ are the hyperradial wave functions, where the subscript $i$ denotes the hyperradial excitation (for the purpose of solving the coupled equations~\eqref{eq:Faddeev_coupled} the hyper-radial functions, $ \mathcal{R}_{\beta}^{j}(\rho) $ are expanded  in terms of orthonormal discrete basis up to $i_{max}$~\cite{face}).
{Moreover}, the $\beta \equiv \{ K,l_x ,l_y ,l,S_x ,j_{ab} \}$ is a set of quantum numbers coupled to $ j $.  $l_x $ and $l_y $ are the orbital angular momenta related to the Jacobi coordinates $ \vec{x} $ and $ \vec{y} $, respectively.  $l=l_x +l_y $ represents the total orbital angular momentum, $S_x$ gives the total spin of pair particles related by $ \vec{x} $, and $j_{ab}=l+S_x$. 
The total angular momentum is $ j = j_{ab}+I $ where $ I $ indicates the spin of the third particle.

Therefore wave function of the system is defined by 
\begin{eqnarray}
	\Psi^{j\mu}(\rho,\phi) & = & \sum_{\beta}\sum_{i=0}^{i_{max}}C_{i\beta}^{j}\:\psi_{i\beta}^{j\mu}(\rho,\phi)\\
	& = & \sum_{\beta}\left(\sum_{i=0}^{i_{max}}C_{i\beta}^{j}R_{i\beta}(\rho)\right)\mathcal{Y}_{\beta}^{j\mu}(\phi)=\sum_{\beta}\mathcal{R}_{\beta}^{j}(\rho)\mathcal{Y}_{\beta}^{j\mu}(\phi),\nonumber 
	\label{eq:i_max}
\end{eqnarray}
where $ C_{i\beta}^{j} $ are the diagonalization coefficients that can be calculated by
diagonalizing the three-body Hamiltonian for $ i = 0, . . . , i_{max} $ basis functions. 
The hyperradial wave functions $\mathcal{R}_{\beta} (\rho )$ are solutions to the coupled set of differential equations,
\begin{equation}
	\left(-\frac{\hbar^{2}}{2m}\left(\frac{d^{2}}{d\rho^{2}}-\frac{(K+3/2)(K+5/2)}{\rho^{2}}\right)-E\right)\mathcal{R}_{\beta}^{j}(\rho)+\sum_{\beta'}V_{\beta'\beta}^{j\mu}(\rho)\mathcal{R}_{\beta'}^{j}(\rho)=0,
	\label{eq:Faddeev_coupled}
\end{equation} 
the term $ V_{\beta'\beta}^{j\mu}(\rho) $ is related to the two-body potentials between each pair
of particles ($V_{{ij}}$), by
\begin{equation}
	V_{\beta'\beta}^{j\mu}(\rho)=\left\langle \mathcal{Y}_{\beta}^{j\mu}(\phi)\left|V_{12}+V_{13}+V_{23}\right|\mathcal{Y}_{\beta^{\prime}}^{j\mu}(\phi)\right\rangle. 
\end{equation}

\section{Two-body potentials}
\label{sec:Two-body-potentials}
In this section, we introduce  the two-body interaction of $ {NN} $ and different analytical form of extracted lattice $ {N-}\phi $ potential which we used in our calculations for $\phi{-NN}$ system.

\subsection{$ {NN} $ potentials}
For $ {NN} $ interactions, we use  the Yukawa-type Malfliet-Tjon (MT)~\cite{MALFLIET1969161},

\begin{equation}
	V_{NN}\left(r\right)=\sum_{i=1}^{2}C_{i}\frac{e^{-\mu_{i}r}}{r}, \label{eq:VNN}
\end{equation}
the parameters  $ C_{i} $ and $ \mu_{i} $ are given in Table~\ref{tab:MT_para}. This potential supports a deuteron binding energy of $-2.2307$ MeV. 
\begin{table}[htp]
	\caption{The parameters and low-energy scattering data of the local central MT $ {NN} $ potential given in Eq.~\eqref{eq:VNN} for the singlet $ ^{1}S_{0} $ and triplet $ ^{3}S_{1} $ channel.}
		\begin{tabular}{ccccccc}
			\hline
			$ \left(I,J\right) $ & $ a_{0}\left(\mathrm{fm}\right) $&$ \mathrm{{r_{\mathrm{eff}}}}\left(\mathrm{fm}\right) $&$C_1~(\mathrm{MeV \cdot fm})$ & $\mu_1~(\mathrm{fm}^{-1})$ & $C_2~(\mathrm{MeV \cdot fm})$ & $\mu_{2}(\mathrm{fm}^{-1})$ \\
			\hline  
			$ \left(1,0\right) $&$ -23.56 $&$ 2.88 $&$ -513.968 $  & $ 1.55 $  & $ 14.38.72 $  &$  3.11 $  \\
			\hline
			$ \left(0,1\right) $&$ 5.51 $   &$ 1.89$&$ -626.885 $  & $ 1.55 $  & $ 1438.72 $  &$  3.11 $  \\
			\hline
		\end{tabular}
		\label{tab:MT_para}
	\end{table}
	
	\subsection{$ {N-}\phi \left(^{4}S_{3/2}\right) $ potential} 
	\label{subsec:Spin-2--potential}
	For the $ {N-}\phi $ potential  in the $ ^{4}S_{3/2} $  channel, i.e.,  the concrete parameterizations, are taken straight from Ref.~\cite{yan2022prd} which is published very recently by the HAL QCD collaboration. 
Where they performed the uncorrelated fit on the lattice QCD extracted potential  to calculate physical observables 
	 using two different analytic functional forms composed of attractive Gaussian and long-range Yukawa squared attractions~\cite{etminan2014,Iritani2019prb}, 
	\begin{equation}
		V_{A}\left(r\right)=\sum_{i=1}^{2}\alpha_{i}e^{-\left(r/\beta_{i}\right)^{2}}+a_{3}m_{\pi}^{4}f\left(r;\beta_{3}\right)\left(\frac{e^{-m_{\pi}r}}{r}\right)^{2}. \label{eq:pot_nphi_A}
	\end{equation}
	In Ref.~\cite{yan2022prd}, it is shown that the long-range part of the $ {N-}\phi $ potential
	is clearly dominated by the two-pion exchange (TPE). This behavior suggests the $ V_{A}\left(r\right) $ {has} a TPE tail at long range with a strength coefficient $ m_{\pi}^{4} $~\cite{PhysRevD.98.014029}. Also, for comparison, a purely phenomenological Gaussian form is considered, 
	\begin{equation}
		V_{B}\left(r\right)=\sum_{i=1}^{3}\alpha_{i}e^{-\left(r/\beta_{i}\right)^{2}}. \label{eq:pot_nphi_b}
	\end{equation}
	For the form factor $ f\left(r;b_{3}\right) $ in Eq.~\eqref{eq:pot_nphi_A} two different {types} commonly used in the $ {NN} $ {potential}, is applied
	a Nijmegen-type form factor~\cite{PhysRevC.49.2950}, 
	\begin{eqnarray}
		f_{{erfc}}\left(r;\beta_{3}\right)=\left[{erfc}\left(\frac{m_\pi}{\Lambda}-\frac{\Lambda r}{2}\right)-{e}^{2m_\pi r}{erfc}\left(\frac{m_\pi}{\Lambda}+\frac{\Lambda r}{2}\right)\right]^{2}/4,
	\end{eqnarray}
	and the Argonne-type form factor~\cite{PhysRevC.51.38},
	\begin{eqnarray}
		f_{{exp}}\left(r;\beta_{3}\right)=\left(1-e^{-(r/\beta_{3})^{2}}\right)^{2},
	\end{eqnarray}
	where the lattice pion mass is $m_{\pi} = 146.4 $ MeV, $ \Lambda=2/\beta_{3} $ and $ {erfc}\left(z\right)=\frac{2}{\sqrt{\pi}}\int_{z}^{\infty}e^{-t^{2}}dt $ is the complementary error function. Hereafter, we refer to $ V_{A}\left(r\right) $ with $ f_{{erfc}}\left({f}_{{exp}}\right) $ form factor as $ {A}_{{erfc}}\left({A}_{{exp}}\right) $ model, and model B is applied to $ V_{B}\left(r\right) $ in Eq.~\eqref{eq:pot_nphi_b}. The parameters of Eq.~\eqref{eq:pot_nphi_A} and Eq.~\eqref{eq:pot_nphi_b} those we employed in our calculations are given in Table~\ref{tab:Fit_para}. 
	The central values of low-energy observables by $ {A}_{{erfc}} $ potential are  the scattering length $a_{0}^{{N-}\phi}=-1.43(23)$ fm, and the effective range $r_{eff}^{{N-}\phi}=2.36(10)$ fm \cite{yan2022prd} and  no binding energy is observed for this interaction. The number between   parentheses is the statistical error.

	\begin{table}[htp]
		\caption{The parameters of  $ {N-}\phi $  $ \left(^{4}S_{3/2}\right) $ potential given in Eqs.~\eqref{eq:pot_nphi_A} and ~\eqref{eq:pot_nphi_b} from Ref.~\cite{yan2022prd} at lattice Euclidean time $ 14 $. The numbers in  parentheses indicate statistical errors.}
			\begin{tabular}{ccccccc}
				\hline
				&$\alpha_1~(\mathrm{MeV})$ &	$\beta_1~(\mathrm{fm})$ &	$\alpha_2~(\mathrm{MeV})$ & $\beta_2~(\mathrm{fm})$ & $\alpha_3 m_{\pi}^{4n} ~(\mathrm{MeV \cdot fm^{2n}})$ & $\beta_{3}(\mathrm{fm})$ \\
				\hline  
				$ {A}_{{erfc}} $&$ -376(20) $  &$ 0.14(1) $  &$ 306(122) $  & $ 0.46(4) $ & $ -95(13) $  & $ 0.41(7) $  \\
				\hline
				$ {A}_{{exp}} $&$ -371(27 $)  &$ 0.13(1) $  &$ -119(39) $  & $ 0.30(5) $ & $ -97(14) $  & $ 0.63(4) $  \\
				\hline
				$ {B-3G} $&$ -371(19 $)  &$ 0.15(3) $  &$ -50(35) $  & $ 0.66(61) $ & $ -31(53) $  & $ 1.09(41) $  \\
				\hline
			\end{tabular}
			\label{tab:Fit_para}
		\end{table}

\begin{figure*}[hbt!]
			\centering
			\includegraphics[scale=1.0]{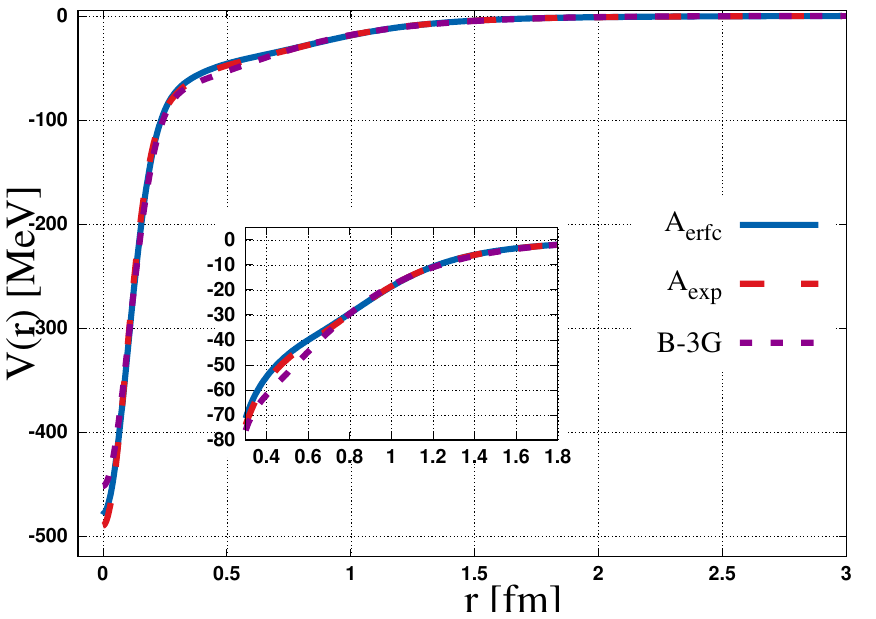}
			\caption{The  $ {N-}\phi $ potential in the $ ^{4}S_{3/2} $ channel as a function of separation $ r $ at  lattice Euclidean time $ 14 $ from Ref.~\cite{yan2022prd}, for three model, i.e. $ {A}_{{erfc}} $ (blue solid line), $ {A}_{{exp}} $ (red dashed lines), and $ {B-3G} $ (purple dot line) using the parameters given in Table~\ref{tab:Fit_para}.}
			\label{fig:Nphi_pot}
		\end{figure*}

		In Fig.~\ref{fig:Nphi_pot} we show all three $ {N-}\phi $ potential models in the $ ^{4}S_{3/2} $ channel using the parameters given in Table~\ref{tab:Fit_para}, for better comparison.
		\section{Numerical Results} 
		\label{result}
		Here,  we present and discuss our numerical results for the ground state binding energy $ B_3 $ and nuclear matter radius $ r_{mat} $ for three-body   $ \phi{-NN} $ state systems. For this purpose, the coupled equations~\eqref{eq:Faddeev_coupled} {are} solved by applying the FaCE computational toolkit~\cite{face} employing the two-body interactions described in Sec.~\ref{sec:Two-body-potentials}.

		In the HH method, the final results  depend on a maximum value of hypermomentum $ K_{max} $ due to the expansion of the total three-body wave function in hypermomentum components, which are truncated by $ K_{max} $, therefore, it is necessary to investigate the convergence of results as a function of $ K_{max} $ and the maximum number of hyperradial excitations $i_{max}$ (see Eq.~\eqref{eq:i_max}). 
		The results converged quite well with the $ K_{max} = 70$ and $ i_{max} =25 $ for $(I)J^{\pi}=(0)2^{-}$ $ \phi{-d} $ state.  
		
		The $ {N-}\phi $ two-body system in $ ^{4}S_{3/2} $ channel(~Fig.~\ref{fig:Nphi_pot}) by parametrizations of all three model ${A}_{{erfc}} $, $ {A}_{{exp}} $, and $ {B-3G} $  does not form a bound state as previously predicted in~\cite{PhysRevD.84.056017,RAMOS2013287} by using the hidden gauge theory with unitary coupled-channel calculations within $ {SU}(3) $ symmetry,  whereas  within chiral $ {SU}(3) $ quark model~\cite{PhysRevC.73.025207} and unitary coupled-channel approach~\cite{Sun_2023}  the existence of a bound state is reported.
		In older studies, as mentioned in the introduction, in Refs.~\cite{Belyaev2008,Belyaev2009,Sofianos2010} by using the attractive  $ {N-}\phi $ interaction~\cite{PhysRevC.63.022201}, it is predicted that the binding energy of $\phi{-NN}$  state could be about $ 40\left(23\right) $ MeV for triplet(singlet) $ {NN} $ interaction.

		Nevertheless, in this case, as  expected no bound state found for $(I)J^{\pi}=(1)1^{-}$ $ \phi{-NN} $ state, i.e. $ \phi{-nn} $ and $ \phi{-pp} $. While the $ (I)J^{\pi}=(0)2^{-}$ $ \phi{-d} $ state in the maximal spin channel is bound for all parametrizations of the $ {N-}\phi $ potentials, i.e. 
		$ {A}_{{erfc}} $, $ {A}_{{exp}} $, and $ {B-3G} $. The ground state binding energy and nuclear matter radius of $ \phi{-d} $ state are given in Table~\ref{tab:pot3}. To calculate the r.m.s. matter radius of the $ \phi{-d} $ system, the  strong interaction radius of proton, neutron and $ \phi $-meson by the values $ 0.82, 0.80 $ fm and $ 0.46 $ fm~\cite{PhysRevC.108.034614,POVH1990653}, respectively, are have been used as input.
		
		Moreover, the results  with the $ m_{N}$ and $ m_{\phi}$ masses derived 
		from $\left(2+1\right)$-flavor lattice QCD simulations~\cite{yan2022prd}, presented in Table~\ref{tab:pot3}.
		The values of these masses {are} slightly bigger than the experimental value. As it can be seen from the Table~\ref{tab:pot3}, $B_3$ by lattice masses are a bit larger than $ B_3 $ by the experimental masses. This is because by increasing the masses,  
		repulsive kinetic energy contribution will decrease  which in turn  leads to an increment in binding energies~\cite{Garcilazo2019}. 
		
		%
		\begin{table}
			\caption{Three-body ground state binding energy $\left(B_{3}\right)$ and the nuclear matter radius $ \left(r_{{mat}}\right) $ of the $(I)J^{\pi}=(0)2^{-}$ $ \phi{-d} $ state for three type $ {N-}\phi $ potentials, i.e.  
				$ {A}_{{erfc}} $, $ {A}_{{exp}} $, and  $ {B-3G} $. The results were calculated using experimental values of masses,
				$m_{N}=938.9$ ${MeV}/{c}^{2}$ and $m_{\phi}=1019.5$
				${MeV}/{c}^{2}$, and with the masses obtained
				from $\left(2+1\right)$-flavor lattice QCD simulations, $m_{N}=954.0$
				${MeV}/{c}^{2}$ and $m_{\phi}=1048.0$ ${MeV}/{c}^{2}$~\cite{yan2022prd}. }
			\centering
				\begin{tabular}{ccccccccc}
					\hline
					&	  \multicolumn{2}{c}{$ {A}_{{erfc}} $}    && \multicolumn{2}{c}{${A}_{{exp}}$}    &&  \multicolumn{2}{c}{${B-3G}$} \\ \cline{2-3} \cline{5-6} \cline{8-9}
					&   $B_3$ (MeV) & $r_{mat}$ (fm) && $B_3$ (MeV) & $r_{mat}$ (fm) && $B_3$ (MeV) & $r_{mat}$ (fm) \\
					\hline  
					Expt.    &  $ 6.9 $ & $ 8.33 $ && $ 6.8 $ & $ 8.24 $ && $ 6.7 $ & $ 8.08 $  \\
					\hline
					Lattice  &  $ 7.3 $ & $ 8.35 $ && $ 7.2 $ & $ 8.25 $ && $ 7.1 $ & $ 8.05 $  \\
					\hline
				\end{tabular}
				\label{tab:pot3}
			\end{table}
			
				\section{Summary and conclusions\label{sec:Summary-and-conclusions}}
				In this work, the binding energy of the three-body system $ \phi-NN $ is examined using the 
				first lattice QCD  $ N-\phi $ potential  in the $ ^{4}S_{3/2} $  channel  and semi-realistic Malfliet-Tjon $ {NN} $ interactions. The $ N-\phi $ potential is obtained from QCD on a sufficiently large lattice at almost physical quark masses ($m_{\pi}\simeq146.4$ MeV
				and $m_{K}\simeq525$ MeV) and parametrized in three different analytical forms, i.e. $ {A}_{{erfc}} $,   $ {A}_{{exp}} $ and $ {B-3G} $, where the concrete parameterizations of these models, are taken straight from Ref.~\cite{yan2022prd}.

								Then, by having a two-body potential between each pair of particles, the coupled Faddeev equations in the coordinate space are solved within the hyperspherical harmonics expansion method.
								
				We have tried to find bound states or resonances that can be observed in future experiments.
				The numerical results suggest that 
				no bound state  or resonances found for $(I)J^{\pi}=(1)1^{-}$  $ \phi{-nn} $ and $ \phi{-pp} $ systems. The $ (I)J^{\pi}=(0)2^{-}$ $ \phi{-d} $ system in the maximal spin  presents  a bound state with a binding energy of about $ 7 $ MeV and a nuclear matter radius of about $ 8 $ fm.
				The $ \phi{-d} $ system in the maximal spin channel
				cannot couple to the lower three-body open channels $\Lambda K N$ and
				$ \Sigma K N $ because   $ D $
				wave subsystems $ \Lambda K $ and $ \Sigma K $ are kinematically suppressed at low energies.
				In addition, because of the small phase space, the decay to
				final states by four or more particles e.g. $\Sigma \pi K N, \Lambda \pi K N$, and
				$ \Lambda \pi \pi K N $ are supposed to be suppressed~\cite{yan2022prd}.
				Last but not least, in Ref.~\cite{yan2022prd}  they have not consider the OZI  violating $ s \bar{s} $ annihilation in their simulations. Nevertheless, considering the coupling to channels
like $ \rho N $ or $ \pi \Delta $ could change the results obtained here significantly.

				In conclusion, adding a $ \phi $-meson to the deuteron could enhance its binding energy. These bound states or resonances could be explored in  hadron beam experiments.
				Recently, for the first time, ALICE
				collaboration measured the correlation function of proton-$\phi$  in heavy-ion collisions~\cite{PhysRevLett.127.172301}, 
				and the  existence  of $ p{-}\phi $ bound state has been discussed and explored in~\cite{CHIZZALI2024138358}. And very recently, the $ K^{+}{-d} $ and $ p{-d} $ femtoscopic correlations measured by the ALICE Collaboration in proton-proton (pp) collisions~\cite{alicecollaboration2023exploring,mrowczynski2019hadron}, analogously, as the next step, these measurements could be done for $ \phi{-d} $ system.  We desire that our numerical results could help to
				plan experiments in the future.

				

				
				\bibliography{Refs.bib}

\end{document}